# Altering sensorimotor simulation impacts early stages of facial expression processing depending on individual differences in alexithymic traits

Arianna Schiano Lomoriello[1,2^], Antonio Maffei[3^], Sabrina Brigadoi[1,4] and Paola Sessa[1,3*]

[1] Department of Developmental and Social Psychology, University of Padova, Padova, Italy

[2] Department of Cognitive System, Denmark Technical University (DTU), Copenhagen, Denmark

[3] Padova Neuroscience Center (PNC), University of Padova, Padova, Italy

[4] Department of Information Engineering, University of Padova, Padova, Italy

^ These authors contributed equally to the work.

\* **Corresponding author:** Paola Sessa, Department of Developmental and Social Psychology, University of Padova, Via Venezia 8, 35121, Padua, Italy.

paola.sessa@unipd.it

**Abstract**

Simulation models of facial expressions suggest that posterior visual areas and brain areas underpinning sensorimotor simulations might interact to improve facial expression processing. According to these models, facial mimicry, a manifestation of sensorimotor simulation, may contribute to the visual processing of facial expressions by influencing early stages. The aim of the present study was to assess whether and how early sensorimotor simulation influences early stages of face processing. A secondary aim was to investigate whether there is a relationship between alexithymic traits and sensorimotor simulation as a mechanism for fine facial expressions discrimination. In order to examine the time-course of face processing, we monitored P1 and N170 components of the event-related potentials (ERP) in participants performing a fine discrimination task of facial expressions while implementing an animal discrimination task as control condition. In half of the experiment, participants could freely use their facial mimicry whereas in the other half, they had their facial mimicry blocked by a gel. Our results revealed that, on average, both P1 and N170 ERP components were not sensitive to mimicry manipulation. However, when taking into account alexithymic traits, a scenario corroborating sensorimotor simulation models emerged, with two dissociable temporal windows affected by mimicry manipulation as a function of alexithymia levels. Specifically, individuals with lower alexithymic traits showed modulations of the P1 amplitude (i.e., larger in the blocked mimicry condition compared to the free mimicry condition), while individuals with higher alexithymic traits showed modulations of the later N170 (i.e., larger amplitude in the blocked mimicry condition compared to the free mimicry condition). Furthermore, connectivity analysis at the scalp level suggested increased connectivity between sensorimotor and extrastriate visual regions in individuals with lower alexithymic traits compared to individuals with higher alexithymic traits. Overall, we interpreted these ERPs modulations as compensative visual processing under conditions of interference on the sensorimotor processing, providing evidence in favor of sensorimotor simulation models.

**Introduction**

Neurobiological models of face processing propose that posterior areas, responsible for the visual analysis of faces, and central and frontal regions, committed to the extraction of emotion and the recovery of semantic and biographical information, interact in order to assign meaning to faces (Calder & Young, 2005; Hoffman, Gobbini, & Haxby, 2000). Among the most endorsed neurobiological models of face processing, the model by Haxby and colleagues (Hoffman, Gobbini, & Haxby, 2002; Hoffman et al., 2000) comprises a core system (including fusiform face area, occipital face area and superior temporal sulcus) for the visual analysis of faces, and an extended system for the advanced processing mentioned above, encompassing a large number of brain regions (including medial prefrontal cortex, temporo-parietal junction, anterior temporal cortex, precuneus, inferior parietal/frontal operculum, intraparietal sulcus, frontal eye fields and the limbic system). An interesting feature that was included in a later revision of Haxby and colleagues' model is "motor simulation" as a mechanism for assigning a meaning to facial expressions and, therefore, for the attribution of emotions (Haxby & Gobbini, 2011).

Models of face and facial expressions processing – including that by Haxby and Gobbini (2011) – do not expand on the exact mechanism by which this simulation process takes place and contributes to emotion recognition and understanding, although several lines of research support the role of simulation in this regard (see, e.g., Gallese & Sinigaglia, 2011; Goldman & de Vignemont, 2009; Pitcher, Garrido, Walsh, & Duchaine, 2008). Neural underpinnings of the motor simulation would include the mirror neuron system (MNS), the premotor cortex (PMC), the inferior parietal lobe (IPL) and the frontal operculum (FO) (see, e.g., Banissy et al., 2011; Montgomery & Haxby, 2008; Montgomery, Seeherman, & Haxby, 2009).

A conceivable manifestation of this motor simulation is the phenomenon known as "facial mimicry" which consists in the visible or non-visible use of facial musculature of an observer matching the other person's facial expression. This phenomenon can be monitored by electromyography (EMG). In support of this hypothesis, several findings provide evidence on how the observer's facial mimicry responds in a congruent fashion to the observed facial expression (e.g., Achaibou, Pourtois, Schwartz, & Vuilleumier, 2008; Dimberg & Petterson, 2000; Korb, Grandjean, & Scherer, 2010) and that a blockage/alteration of facial mimicry reduces the ability to recognize/discriminate facial expressions, both in healthy subjects (Baumeister, Papa, & Foroni, 2016; Baumeister, Rumiati, & Foroni, 2015; Niedenthal, Brauer, Halberstadt, & Innes-Ker, 2001; Oberman, Winkielman, & Ramachandran, 2007; Rychlowska et al., 2014; Stel & van Knippenberg, 2008; Wood, Lupyan, Sherrin, & Niedenthal, 2016) and in neurological patients with partial/total mimic inability, as in the case of patients with facial paralysis (Keillor, Barrett, Crucian, Kortenkamp Sarah, & Heilman, 2002; Korb et al., 2016).

To note, additional brain areas may be implicated in this simulation mechanism, especially the somatosensory cortex (SC), and in this regard the term "sensorimotor simulation" seems more appropriate (e.g., Wood, Rychlowska, Korb, & Niedenthal, 2016). A meta-analysis conducted on patients with focal brain lesions, revealed that damage to the right SC is associated with deficits in the recognition of observed expressions (Adolphs, Damasio, Tranel, Cooper, & Damasio, 2000). Studies that used transcranial magnetic stimulation of the right SC demonstrated the critical involvement of this region in the processing of others' emotions (Adolphs et al., 1999; Hussey & Safford, 2009; Pitcher et al., 2008; Pourtois et al., 2004) and, crucially for the purposes of the present investigation, revealed the sequential involvement of extrastriate areas (60-100 ms) and right SC (100-170 ms) in facial expression recognition (Pitcher et al., 2008). These results support the hypothesis that sensorimotor simulation can influence early stages of face processing. Although previous evidence indicates that regions underpinning sensorimotor simulation and regions responsible for visual analysis of faces interact, it is still unclear how early this interaction occurs during face and facial expression processing. A recent study by Sessa and colleagues (Sessa, Schiano Lomoriello, & Luria, 2018) showed that visual working memory (VWM) representations of faces are affected by the blockage/alteration of the observers' facial mimicry. In a change detection task (Luria, Sessa, Gotler, Jolicœur, & Dell'Acqua, 2010; Meconi, Luria, & Sessa, 2014; Sessa & Dalmaso, 2016; Sessa, Luria, Gotler, Jolicœur, & Dell'Acqua, 2011; Sessa et al., 2012; Vogel & Machizawa, 2004; Vogel, McCollough, & Machizawa, 2005), participants had to memorize a pre-cued and lateralized face expressing a certain intensity of anger (memory array) and decide, about 1 second later, whether a second face (test array) presented in the same hemifield and of the same individual had the same or a different intensity of facial expression. Critically, participants could freely use their facial mimicry whilst performing the task in half of the experiment whereas in the other half their mimicry was altered/blocked by a hardening facial gel (with the order of the two conditions counterbalanced across participants). In a 300-1300 ms time-window between the memory and the test array, a component of the event-related potentials (ERPs) indexing the quantity/quality of VWM representation was monitored. The amplitude of this ERP component, known as sustained posterior contralateral negativity (SPCN; Jolicœur et al., 2007; Luria et al., 2010; Meconi et al., 2014; Sessa and Dalmaso, 2016; Sessa et al., 2011, 2012) or as contralateral delay activity (CDA; Vogel and Machizawa, 2004) was found to be reduced in the condition in which participants' mimicry was blocked/altered when compared to the condition in which they could freely use their mimicry. Furthermore, participants who were most affected by the mimicry manipulation were the most empathic on the basis of their scores in the Empathy Quotient questionnaire (EQ; Baron-Cohen & Wheelwright, 2004), in line with the evidence suggesting that the most empathic individuals are those using their facial mimicry more and they are also characterized by a greater susceptibility to emotional contagion (Balconi & Canavesio, 2016; Bos, Jap-Tjong, Spencer, & Hofman, 2016; Dimberg, Andréasson, & Thunberg, 2011; Prochazkova & Kret, 2017; Seibt, Mühlberger, Likowski, & Weyers, 2015; Sonnby-Borgström, 2002; but see also Franzen, Mader, & Winter, 2018).

Thus, this latter evidence suggests that a high-level visual processing stage, i.e. VWM, may be influenced by the sensorimotor simulation activity during face/facial expression processing. The model by Wood and colleagues (2016) proposes that these effects might be observed even earlier during initial stages of face processing, possibly already at the stage of faces structural encoding (George, Evans, Fiori, Davidoff, & Renault, 1996; Jeffreys, 1983; Perez, Mccarthy, Bentin, Allison, & Puce, 1996). Furthermore, along the line of Pitcher and colleagues' findings (2008), which revealed somatosensory activity 100-170 ms after the presentation of facial expressions, this hypothesis seems even more plausible.

The aim of the present study was exactly to provide a direct test of the hypothesis that sensorimotor activity can influence early stages of face processing. The recent simulation model by Wood and colleagues (2016) hypothesizes indeed an iterative process between the posterior regions of visual processing and the sensorimotor regions. Therefore, it is reasonable to hypothesize that the effects of sensorimotor simulation on face visual processing should be considered within a cascade model, in which the impact of sensorimotor activity possibly becomes increasingly evident during processing in the extrastriate areas.

ERP responses to faces can be observed as early as 100 ms following their presentation, as indicated by modulations of the P1 ERP component (Herrmann, Ehlis, Ellgring, & Fallgatter, 2005; Marzi & Viggiano, 2007); the face-sensitive P1 ERP response is followed by another ERP that maps the structural encoding stage, i.e. the N170 (Perez et al., 1996), an occipitotemporal response characterized by a negative polarity with a peak latency of approximately 170 ms and the largest amplitude to faces. The N170 has been observed both for photographic face images and for face drawings (Perez et al., 1996; Sagiv & Bentin, 2001) and it is sensitive to the inversion effect typically associated with larger N170 amplitude, and often increased latency, for inverted when compared to upright faces (e.g., Anaki, Zion-Golumbic, & Bentin, 2007; Caharel, Fiori, Bernard, Lalonde, &

Rebaï, 2006; Eimer, 2000a; Itier, Latinus, & Taylor, 2006; Jacques & Rossion, 2007; Marzi & Viggiano, 2007; Perez et al., 1996; Rossion, Delvenne, Debatisse, & Goffaux, 1999). On the contrary, it is generally believed to be insensitive to face familiarity and identity (e.g., Bentin & Deouell, 2000; Eimer, 2000). It probably arises at the scalp level as the activation of multiple cortical sources including the fusiform face area (see, e.g., Mnatsakanian & Tarkka, 2004), the occipital face area (see, e.g., Deffke et al., 2007) and the posterior part of the superior temporal sulcus (Watanabe, Kakigi, & Puce, 2003), all of these regions being compatible with the "core system" of the model by Haxby and colleagues (Hoffman et al., 2000).

In the present investigation we employed the ERP technique in a within-subjects design. We administered our participants a task similar to that by Wood et al. (2016). This previous behavioral study – conducted on a large sample of participants (N = 122) – involved, in a between-subjects design, a mimicry manipulation by means of a facial gel able to block/alter the participants' facial mimicry during a task that involved distinguishing target expressions from highly similar distractors. Stimuli could be both faces, selected from a morphing continuum of a face identity from an expression of 100% anger to an expression of 100% sadness, and animals selected from a morphing continuum from the image of a horse (100%) to the image of a cow (100%), as a control condition. The results showed that blocking/altering facial mimicry had a selective negative impact on the accurate discrimination of facial expressions. The authors then proposed that this decrease in accuracy in the fine discrimination of emotions was due to a selective interference with the simulation process that in turn would not have contributed (or would have to a small extent) to the construction of face visual percepts. Although exciting, this evidence is indirect and does not allow to reach these intriguing conclusions with certainty.

We employed Wood's et al. (2016) stimuli and then manipulated participants' facial mimicry in a within-subjects design such that participants performed the discrimination task with a hardening facial gel in half of the experiment (with counterbalanced order). By means of ERPs we were able to trace the time course of the effects of mimicry on fine facial expressions discrimination focusing on early components of ERPs associated with face and facial expressions processing, i.e. P1 and N170 ERP components.

We hypothesized that blocking/altering facial mimicry would have affected sensorimotor simulation causing a cascading effect on early face processing and translating into modulations of the P1 and/or N170 ERP components.

A secondary aim of the present study was to start an exploration of the relationship between alexithymic traits and sensorimotor simulation as a mechanism for fine facial expression discrimination. To this purpose, participants completed the Toronto Alexithymia Scale (TAS-20; Bagby, Parker, & Taylor, 1994; Caretti, La Barbera, & Craparo, 2005, for the Italian version) at the end of the experimental electroencephalographic session. Recent studies have suggested that alexithymia – defined as the difficulty of identifying one's own and others' emotions – could be characterized by a deficit in sensorimotor simulation (or embodied simulation; e.g., Gallese & Sinigaglia, 2011; see, e.g., Scarpazza & di Pellegrino, 2018; Scarpazza, Làdavas, & Cattaneo, 2018) during the processing of facial expressions of others' emotions, especially with regard to those with negative valence (Scarpazza, di Pellegrino, & Làdavas, 2014; Scarpazza, Làdavas, & Di Pellegrino, 2015; Sonnby-Borgström, 2009). One of these studies, in particular, demonstrated in alexithymic participants a reduced activity of the corrugator supercilii and of the zygomaticus major, respectively for negative and positive emotions, during the passive view of facial expressions (Sonnby-Borgström, 2009). These previous studies indicate that individuals with greater alexithymic traits might be less – or differently – affected by mimicry manipulations in fine emotions discrimination tasks precisely because they tend to use to a lesser degree the sensorimotor simulation mechanism to recognize and discriminate emotions in others. With regard to the present investigation, we hypothesized to observe a relationship between alexithymic traits and modulations of P1 and/or N170 ERP components as well as accuracy in the fine discrimination task as a function of the mimicry manipulation.

Finally, we also investigated the effect of altering the facial mimicry on the connectivity between visual and sensorimotor regions. According to Wood and colleagues' model (2016), facial expressions' processing occurs within a continuous information exchange between the visual and sensorimotor areas. Thus, another aim of the present investigation was to test this hypothesis by studying whether – and eventually how – this information flow could be affected by altering/blocking facial mimicry, also in relation to alexithymic traits.

**Method**

Participants

Data were collected from 35 volunteer healthy students (6 males) from the University of Padova. Data from two participants were discarded from analyses due to excessive electrophysiological artifacts. All participants included in the final sample reported normal or corrected-to-normal vision and no history of neurological disorders. The final sample included 33 participants (mean age: 22.8 years, SD = 3.28, 4 left-handed) in line with a reference study for this investigation (Sessa et al., 2018; see also Achaibou et al., 2008). All participants signed a consent form according to the ethical principles approved by the University of Padova (Protocol number: 1986).

Stimuli

The stimuli were 11 grayscale digital photographs (i.e., faces and animals stimuli) for each morph *continuum*. We adopted the stimuli developed by Niedenthal and colleagues (Niedenthal, Halberstadt, Margolin, & Innes-Ker, 2000) and then used in Wood and colleagues' experiment (2016). In particular, the face stimuli consisted of images of a female model expressing morphed combinations of sadness and anger emotions, while the non-face control images were selected from a morph of a horse and a cow that had maximally similar postures. Specifically, the face continuum began at 100% sad and 0% angry and transitioned in 10% increments to 0% sad and 100% angry (see Figure 1). All images were resized to subtend a visual angle between 10 and 12 deg. Participants were seated about 60 cm away from the screen.

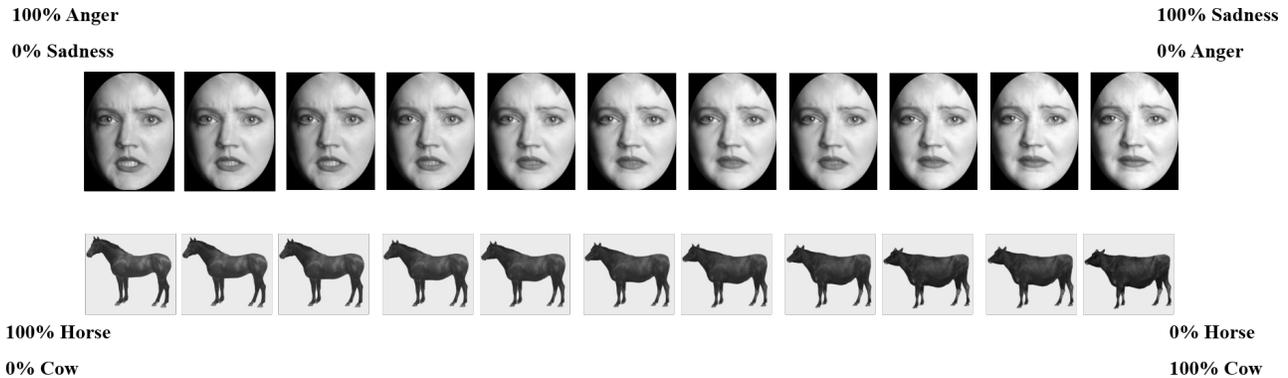

*Figure 1. Stimuli used in the XAB task, for both the facial expressions discrimination and the animal shapes discrimination conditions.*

Procedure

The XAB discrimination task required participants to discriminate a target from a perceptually similar distractor. Before starting the experiment, participants performed twelve practice trials to get familiar with the task. Each trial (Figure 2 depicts the trial structure of the XAB discrimination task) began with a 500 ms fixation cross, followed by the target image (X) for 750 ms. The target was then followed by a 350 ms noise mask, aiming to limit the processing of the stimuli, thus controlling for the potential effects of iconic memory representations. Every trial was interleaved by a variable blank interval (Inter-stimulus Interval, ISI: 800-900 ms). The target image reappeared alongside a distractor, with left–right locations counterbalanced across trials. The target and distractor images could be at 20% apart on the morph continuum, yielding nine image pairs, or 40% apart, yielding six pairs. The motivation for this experimental manipulation is that previous work suggested that sensorimotor simulation may be especially recruited in the case of subtle discrimination of facial expressions (Rychlowska et al., 2014; Wood, Lupyan, et al., 2016; Wood, Rychlowska, et al., 2016). The target and distractor remained on the screen until participants' response. Participants' task was to press a key (F = left; J = right) to indicate which image matched the target image seen in the first screen presentation. Participants performed 4 experimental blocks, two for each level of discrimination (20% or 40% distance between the target and distractor, alternately), with counterbalanced order across participants. Each block consisted of 144 trials (i.e., 576 trials in total).

Each participant performed the task in two different conditions (counterbalanced order across participants); in one, (blocked/altered mimicry condition) a mask gel was applied on the participant's whole face, so as to create a thick and uniform layer, excluding the areas near the eyes and upper lip. The product used as a gel was a removable cosmetic mask (BlackMask Gabrini©) that dries in 10 minutes from application and becomes a sort of plasticized and rigid mask. Participants perceived that the gel prevented the wider movements of face muscles. In the other half of the experiment (free mimicry condition) nothing was applied to the participants' faces.

As in the study by Wood and colleagues (Wood et al., 2016), at the beginning of the experimental session participants were told that the experiment concerned "the role of skin conductance in perception" and that they would be asked to spread a gel on their face in order to "block skin conductance" before completing a computer task.

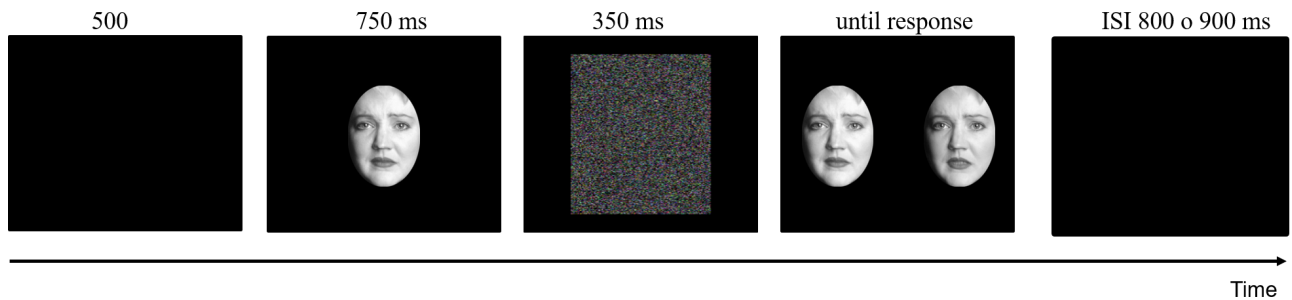

*Figure 2. Timeline of each trial of the XAB task.*

EEG Data Preprocessing

The EEG was recorded during the task by means of 62 active electrodes distributed on the scalp according to the extended 10/20 system, positioning an elastic Acti-Cap with reference to the left ear lobe. Sampling rate was 1000 Hz and the high viscosity of the gel used allowed the impedance to be kept below 10 KΩ. Continuous data were downsampled to 500 Hz, high-pass filtered at 0.1 Hz, re-referenced to the average of all channels and segmented in epochs starting from -500 ms to 1000 ms with respect to stimulus onset. Independent component analysis (ICA) was applied to the segmented data in order to identify and manually remove artifactual activity related to eye-blinks and saccades (Jung et al., 2000).

**Statistical Analysis**

Behavior

A repeated measures analysis was performed *via* logistic mixed effect models using restricted maximum likelihood (REML) estimation to estimate the effect of type of stimulus (faces vs. animals), mimicry (free vs. blocked/altered mimicry) and level of discrimination (20% vs. 40% apart the morph continuum), as fixed effect, on participant's accuracy level. We included participants' variability as random effect, as suggested by Baayen and colleagues (Baayen, Davidson, & Bates, 2008). Then, to discover the relationship between alexithymic traits and sensorimotor simulation, we divided the sample into two groups, based on the median TAS-20 value: low alexithymic scores (TAS-20 ≤ 43; N = 16) and high-scorers (TAS-20 > 50; N = 17). Thus, in a second model we also included the group (high vs. low alexithymic traits) as fixed effect, and its potential interaction with the other factors, to estimate its potential impact on participant's accuracy level. The behavioral analyses were performed using the software R (2.13) with the lmer function from the lme4 package (Bates, Mächler, Bolker, & Walker, 2015).

ERP

For the quantification of the ERPs, data cleaned with ICA were further segmented into epochs starting from -200 ms to 600 ms with respect to stimulus onset, baseline corrected and low pass filtered at 30 Hz. Epochs with a peak-to-peak amplitude exceeding ± 50 µV in any channel were identified using a moving window procedure (window size = 200 ms, step size = 50 ms) and discarded from further analysis.

For each experimental condition, P1 and N170 activities were defined as the mean amplitude of each participants' grand average in the time windows 100-120 ms and 160-180 ms, respectively, and averaged in two ROIs, one per each hemisphere, clustering together the activity of channels PO7-PO9-O1 (left hemisphere) and PO8-PO10-O2 (right hemisphere). In order to test the effect of wearing the gel mask, a delta score was derived subtracting the activity in the free-mimicry condition from the one in the blocked/altered mimicry condition, i.e. *ΔP1* and *ΔN170*, which represented the units of observation for statistical analysis.

Statistical analysis of ERP activity was performed using linear mixed-effects model adopting a model selection strategy based on the Akaike Information Criterion (*AIC*) (Maffei, Spironelli, & Angrilli, 2019; Wagenmakers & Farrell, 2004). *AIC* (Akaike, 1973) is a powerful metric derived from Information Theory which, starting from a set of candidate models, allows to derive the relative quality of each model (the lowest the *AIC* the highest the quality of the model, controlling for its complexity). For each ERP component, data were fitted with a full model including four fixed-effect predictors (type of stimulus, level of discrimination, hemisphere and group), their interactions and a random intercept to model repeated measurements across subjects. This model was compared with its simpler instances by removing the predictors until reaching an intercept-only model. The best model with lowest *AIC* was identified and the significance of its predictors was assessed with an *F*-test using the Satterthwaite approximation for degrees of freedom (Luke, 2016). In addition, significant effects have been explored using post-hoc pairwise contrasts, corrected for multiple comparisons using false discovery rate (*FDR*; Benjamini and Hockberg, 1995).

Connectivity

In order to characterize the dynamic information flow between visual and sensorimotor areas, the instantaneous phase locking value (iPLV) was used. The PLV is a metric that describes the absolute value of the mean phase difference between two signals (Lachaux, Rodriguez, Martinerie, & Varela, 1999), and has been widely used to investigate brain functional connectivity (Sakkalis, 2011). Instantaneous phase synchrony in the beta band (13-30 Hz) between each pair of channels was computed for each experimental condition. Connectivity in the beta band was studied according to evidences suggesting that processing of emotional facial expression modulates oscillatory activity in this spectral range, and that the extent of this modulation is related to individual differences in empathic abilities (Cooper, Simpson, Till, Simmons, & Puzzo, 2013). Furthermore, beta band connectivity during emotional face processing is reduced in individuals with autism compared to typical developing participants (Leung, Ye, Wong, Taylor, & Doesburg, 2014). Finally, there are converging evidences that the core activity of the sensorimotor system is encoded in beta oscillations (Jensen et al., 2005). The average PLV between two ROIs, one over the occipital region (channels: PO7, PO8, PO9, PO10, O1 and O2) and one over the central region of the scalp (channels: C1, C2, C3, C4, C5 and C6), was computed in the same time windows used for ERP analysis (100-120 ms and 160-180 ms), in order to model the dynamic changes in the functional connectivity between visual and sensorimotor cortices.

The statistical approach to test our predictions regarding connectivity was the same employed for ERPs analysis. A linear mixed-effects model including as fixed-effects the predictors condition (free vs. blocked mimicry) and group (high vs. low alexithymic traits) and their interaction and a random intercept for each subject was fitted, separately, to the data for each time window (P1 and N170) and type of stimulus (face and animal). A model selection approach was used to identify the best model explaining the data according to the *AIC*. The significance of the predictors included in the best model was then assessed with an *F*-test using the Satterthwaite approximation for degrees of freedom (Luke, 2016), and significant effects have been explored using post-hoc pairwise contrasts, corrected for multiple comparisons using false discovery rate (FDR).

**Results**

Behavior

Participants' accuracy was higher when the target and distractor were 40% apart on the morph continuum ($\mu = .961$) than 20% ($\mu = .759$). Moreover, they were more accurate with face ($\mu = .870$) than animal stimuli ($\mu = .848$) and when the target and distractor were 40% apart on the morph continuum ($\mu = .9629$) than 20% ($\mu = .759$). No significant differences emerged depending on the mimicry condition; participants were equally accurate in the free ($\mu = .863$) and altered/blocked mimicry condition ($\mu = .856$).

In the second model, we tested the effect of alexithymic traits in participants' accuracy level; there was no differences between individuals who had high ($\mu = .851$) and low alexithymic traits ($\mu = .867$). None interactions reached significance (min *p* = .082*)*.

|  | Estimate | SE | p value |
|---|---|---|---|
| **Dependent variable:** Participants' accuracy | | | |
| Type of stimuli | .019 | .010 | < .001 |
| Mimicry | -.060 | .072 | .40 |
| Level of discrimination | .984 | .106 | < .001 |
| Alexithymic traits | 1.12 | .150 | .41 |

*Table 1. Summary of the behavioral results.*

ERP: *ΔP1*

Models comparison showed that the best model explaining the data observed for the *ΔP1* was the one including as fixed-effects level of discrimination, type of stimulus, group and the interaction between type of stimulus and group ($AIC = 807.9$, $logL = -396.74$, $\Delta AIC^1 = 23.2$). The *F*-test revealed a significant interaction between type of stimulus and group ($F(1,233) = 13.96$, $p < 0.001$). Pairwise comparisons performed on the interaction revealed that, when the facial mimicry was blocked (Figure 3A), *ΔP1* for facial expressions was higher in the participants with low alexithymia traits compared to *ΔP1* for animals ($t(233) = 3.39$, $p = 0.004$), and higher compared to *ΔP1* for facial expressions in the group with high alexithymia traits ($t(44) = 2.86$, $p = 0.01$).

ERP: *ΔN170*

The best model identified for the analysis of *ΔN170* was that including as fixed-effects type of stimulus, group and their interaction ($AIC = 926$, $logL = -456.82$, $\Delta AIC^1 = 17.2$). The *F*-test revealed a significant main effect of stimulus type ($F(1,232) = 4.2$, $p = 0.041$) and a significant interaction between stimulus type and group ($F(1,232) = 4.21$, $p = 0.041$). Pairwise comparisons performed on the interaction revealed that, with facial mimicry blocked by the gel mask (Figure 3B), *ΔN170* for facial expressions was significantly more negative than the one for animals only in the participants with high alexithymia traits ($t(232) = 3.01$, $p = 0.01$).

---

[1] The $\Delta AIC$ was computed as the difference in *AIC* between the best ranked model and the null model, representing a measure of the difference of the quality between the two models.

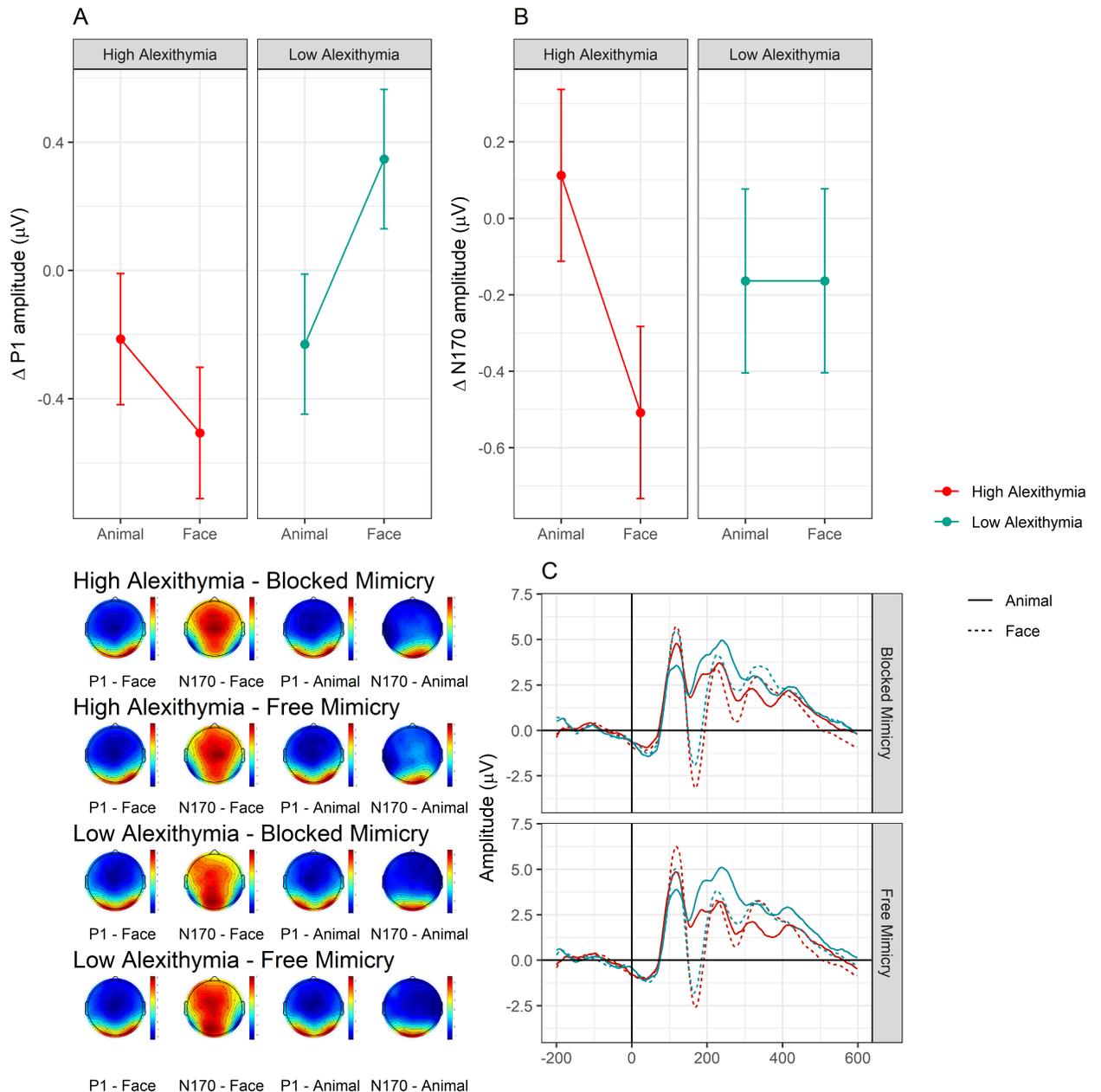

*Figure 3.* Results of ERP analysis. Panel A shows the marginal means for the significant interaction between Type of stimulus and Group for the P1 component. Panel B shows the marginal mean for *significant interaction between Type of stimulus and Group for the N170 component. Panel C shows the target-locked ERP waveforms and the scalp topographies separately for stimulus type, condition and group.*

Phase synchrony: Face stimuli

The best models describing the connectivity in the beta band during the processing of the facial expressions were the full models, including mimicry condition and group and their interaction for both the P1 ($AIC$ = -694.2, $logL$ = 353.28, $\Delta AIC$ = 4.55) and N170 ($AIC$ = -692.4, $logL$ = 352.37, $\Delta AIC$ = 5.27) time windows.

The $F$-test performed on the best model for the P1 time window showed a significant interaction between condition and group ($F(1,229) = 4.85, p = 0.02$). Pairwise comparisons revealed that the connectivity between visual and sensorimotor regions in participants with low alexithymic traits was increased when their mimicry was blocked by the gel mask compared to the free-mimicry condition ($t(229) = 2.49, p = 0.03$), and larger compared to the connectivity observed in those with high alexithymic traits ($t(229) = 2.65, p = 0.03$).

For what concerns the N170 time window, the $F$-test showed a significant group main effect ($F(1,31) = 4.33, p = 0.04$), and a significant interaction between condition and group ($F(1,229) = 5.13, p = 0.02$). Pairwise comparisons performed on the interaction revealed that blocking the facial mimicry elicited a larger connectivity strength in the low alexithymic compared to the participants with high alexithymic traits ($t(46) = 2.84, p = 0.03$).

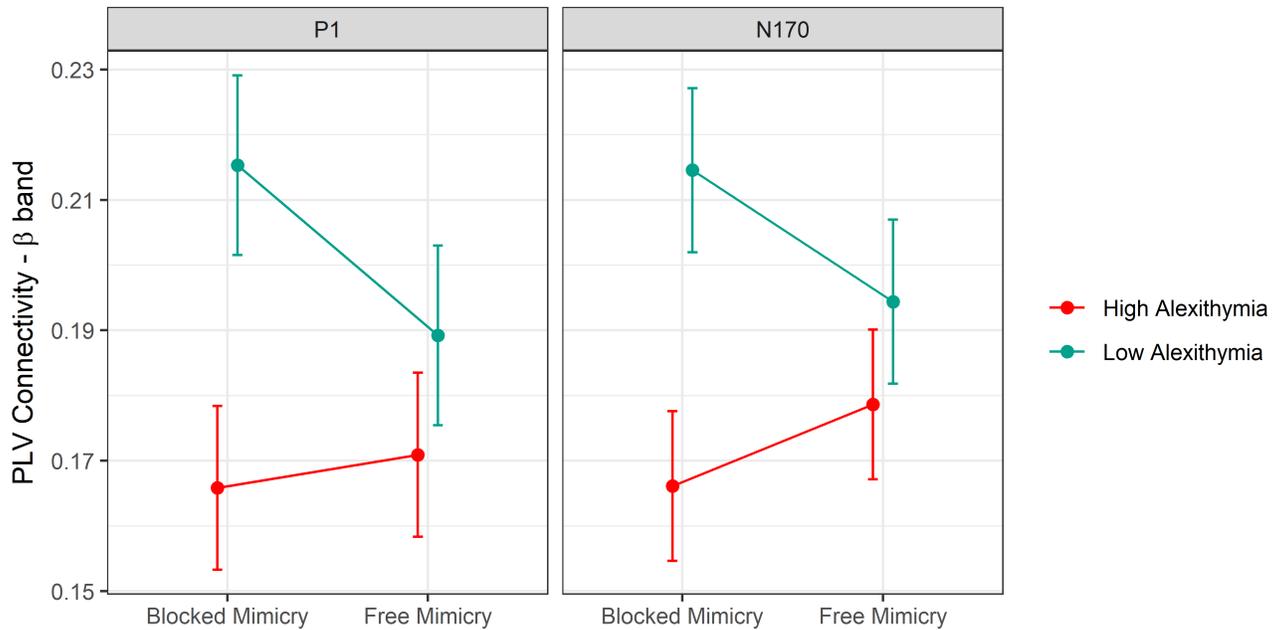

*Figure 4. Results of the connectivity analysis. Marginal means for the significant interaction between Condition and Group during face processing in the P1 time window (left panel) and N170 time window (right panel).*

Phase synchrony: Animal stimuli

The best models describing the connectivity in the beta band during the processing of the animal stimuli were the null model for the P1 time window (*AIC* = -728.7, *logL* = 367.41, *ΔAIC* = 0.23), and the one with only the main effect of mimicry condition and group without interaction for the N170 time window (*AIC* = -759.6, *logL* = 384.92, *ΔAIC* = 5.11).

**Discussion**

By means of the ERP technique, the present investigation had the main objective to monitor the effects of blocking observers' facial mimicry while engaged in a fine facial expression discrimination task (see Wood, et al., 2016) on early stages of face and facial expression processing, reflected in potential modulations of the P1 and N170 ERP components. To assess the selectivity of this effect, a control condition was implemented in which participants had to perform a similar task of fine discrimination of animal shapes (see Wood, et al., 2016). It is important to underline that for this purpose we used a within-subjects manipulation of facial mimicry, so that participants performed one half of the experiment being able to freely use their facial mimicry and a second half of the experiment with their mimicry blocked by a hardening gel (the order of the two conditions was counterbalanced across participants).

The model proposed by Wood and colleagues (2016) suggests an iterative connection between the areas responsible for the visual processing of faces and the sensorimotor areas in charge of simulation processes. Previous evidence also suggests that sensorimotor activity can be observed as early as 100-170 ms from the presentation of facial expressions (Pitcher et al., 2008). Starting from these theoretical and experimental foundations, we first of all hypothesized that early ERP components, i.e., P1 and/or N170, would have been modulated as a function of mimicry manipulation in fine discrimination of facial expressions.

Additionally, it is important to consider that facial mimicry is mainly conceived as a manifestation of sensorimotor simulation and its feedback to the motor and somatosensory regions (*via* the motor regions) seems to be relevant for the dynamic modeling of the simulation itself. In fact, several previous studies have shown how an alteration of facial mimicry is detrimental for the recognition/discrimination of facial expressions (Baumeister et al., 2016; Baumeister et al., 2015; Keillor et al., 2002; Korb et al., 2016; Niedenthal et al., 2001; Oberman et al., 2007; Rychlowska et al., 2014; Stel & van Knippenberg, 2008; Wood, et al., 2016). Based on these evidences, and in line with the theoretical framework proposed by Wood and colleagues (2016), we expected to observe that blocking mimicry would impact connectivity at the scalp level between temporo-occipital and central electrodes selectively for faces (but not for animal stimuli) in the temporal window including the P1 and N170 components.

The present results are in line with the predictions, but reveal a more complex picture than originally expected. Indeed, our results revealed that both the P1 and the N170 ERP components were not significantly modulated as a function of the mimicry manipulation by itself, but only when individual differences in alexithymia are taken in account. Additionally, connectivity analysis reveals that during face processing our manipulation affected the information flow between visual and sensorimotor regions differently in the two groups of participants. Along the same line, ERPs results showed, only in participants with higher alexithymic traits, a greater N170 amplitude values selectively for face stimuli when those wore the facial gel compared to when their mimicry was not blocked/altered. Interestingly, a similar modulation was observed in participants with lower alexithymic traits at the level of the earlier P1 component, i.e. a greater P1 amplitude values selectively for face stimuli when those wore the facial gel compared to when their mimicry was not blocked/altered.

Whilst, on the one hand, we did not expect a modulation of the N170 component in participants with higher alexithymic traits as a function of the mimicry condition, on the other hand, the results at the connectivity analysis shows a reduction of the phase synchrony in the beta band between occipito-temporal and central regions only in participants with higher alexithymic traits. These results are very nicely in agreement with our hypothesis and previous experimental evidence, in which a deficient sensorimotor simulation in alexithymic subjects has been documented, likely arising from abnormal interoceptive abilities (Scarpazza & di Pellegrino, 2018; Scarpazza et al., 2018; Scarpazza et al., 2014, 2015; Sonnby-Borgström, 2009). Additionally, we found exactly the opposite pattern for those individuals with low alexithymic traits where the alteration/blockage of the facial mimicry prompts an increase in the visual-sensorimotor connectivity.

Therefore, combining together these findings provides supports to the hypothesis that a manipulation of facial mimicry has an impact on early stages of face processing, and reveals that it is temporally dissociable according to the level of alexithymia of the subjects. This temporal dissociation likely results from the different degree of connectivity between sensorimotor and visual regions in the two groups, as indicated by the instantaneous phase synchrony in the beta band. Therefore, both modulations, of P1 (in participants with lower alexithymic traits) and N170 (in participants with higher alexithymic traits) would be the consequence of interference on the simulation system induced by the manipulation of facial mimicry. This interpretation is corroborated by the observation that these modulation effects have been selectively observed for face stimuli but not for animal stimuli. It remains unclear whether these modulations reflect disruption of face processing or rather a compensatory mechanism. In this perspective, the null behavioral results in terms of accuracy as a function of the mimicry conditions allow us to reach out for the second interpretation. This whole pattern of findings finds support in more recent evidence. A study by de la Rosa and colleagues (de la Rosa, Fademrecht, Bülthoff, Giese, & Curio, 2018) has recently supported the existence of the two pathways of facial expressions processing, i.e. visual and sensorimotor, sensitive in opposed directions to visual and motor adaptation, thus suggesting that the two systems are dissociable; furthermore, the recent study by Sessa and colleagues (2018) demonstrated that facial mimicry can impact high-level visual processing stages of facial expressions (in terms of modulations of the SPNC ERP component; also named CDA). Overall, this evidence sustains the conclusion that the two systems, visual and sensorimotor, are dissociable but interact, exactly as suggested by the Wood and colleagues' model (2016). In this vein, the impairment of the functioning of one system (e.g., sensorimotor, as in the case of the mimicry manipulation) can be compensated by the other system (e.g., visual). In this perspective, in the absence of appropriate feedback from the sensorimotor regions, greater compensative activity at the level of the regions of the core system (Haxby, 2011) could give rise to P1/N170 of greater amplitude in the condition of blocked mimicry. Therefore, the modulation of the P1/N170 as a function of facial mimicry manipulation would be an expression of a mechanism of visual compensation acting at early stages of visual processing;

furthermore, how early it appears to be a function of the level of alexithymia and the level of connectivity between visual and sensorimotor systems. This compensative mechanism could very clearly explain the absence of a behavioral effect as a function of the facial mimicry manipulation. Finally, these results would fit in part with the EEG-EMG co-registration study by Achaibou, Pourtois, Schwartz & Vuilleumier (2008) who observed that increased levels of facial muscle activity for happy and angry faces were associated with smaller N170 amplitudes, and the authors themselves, in the discussion of their results, suggest that this pattern might result from the existence of two dissociable systems able to compensate each other.

In conclusion, the present study demonstrates for the first time that facial mimicry, as a manifestation of sensorimotor simulation, is able to influence the visual processing of facial expressions at early stages, in particular at the level of the P1 and N170 ERP components. Furthermore, the present results highlight how these modulatory effects are related to individual alexithymic traits, and, in general, to the connectivity between the visual and sensorimotor systems. It is necessary to take into consideration that the participants of our study with higher levels of alexithymia cannot however be classified as alexithymic (except for 2 participants) on the basis of the TAS-20 cut-off scores. It is therefore perhaps legitimate to hypothesize that as the alexithymic traits increase, the sensorimotor activity and/or the connectivity between the sensorimotor and the visual systems is reduced, thus not allowing the implementation by the visual system of the alexithymic subjects of a compensation mechanism. However, this is currently a speculative hypothesis to which future studies will be able to answer more precisely.

Expression, But Not Identity Recognition, in Mirror-Touch Synesthesia. *Journal of Neuroscience*, *31*(5), 1820–1824. https://doi.org/10.1523/JNEUROSCI.5759-09.2011

Baron-cohen, S., & Wheelwright, S. (2004). EQ-an investigation of adults with AS or HFautism and normal sex differences. *Journal of Autism and Developmental Disorder*, *34*(2). https://doi.org/10.1023/B:JADD.0000022607.19833.00

Bates, D., Mächler, M., Bolker, B., & Walker, S. (2015). Fitting Linear Mixed-Effects Models Using lme4. *Journal of Statistical Software; Vol 1, Issue 1 (2015)* . https://doi.org/10.18637/jss.v067.i01

Baumeister, J. C., Papa, G., & Foroni, F. (2016). Deeper than skin deep - The effect of botulinum toxin-A on emotion processing. *Toxicon*, *118*, 86–90. https://doi.org/10.1016/j.toxicon.2016.04.044

Baumeister, J. C., Rumiati, R. I., & Foroni, F. (2015). When the mask falls: The role of facial motor resonance in memory for emotional language. *Acta Psychologica*, *155*, 29–36. https://doi.org/10.1016/j.actpsy.2014.11.012

Benjamini Y. & Hochberg Y. Controlling the false discovery rate: a practical and powerful approach to multiple testing. *Journal of the Royal Statistical Society: Series B*, 57, 298–300 (1995).

Bentin, S., & Deouell, L. Y. (2000). Structural Encoding and Identification in Face Processing: Erp Evidence for Separate Mechanisms. *Cognitive Neuropsychology*, *17*(1–3), 35–55. https://doi.org/10.1080/026432900380472

Bos, P. A., Jap-Tjong, N., Spencer, H., & Hofman, D. (2016). Social context modulates facial imitation of children's emotional expressions. *PLoS ONE*, *11*(12), 1–11. https://doi.org/10.1371/journal.pone.0167991

Caharel, S., Fiori, N., Bernard, C., Lalonde, R., & Rebaï, M. (2006). The effects of inversion and eye displacements of familiar and unknown faces on early and late-stage ERPs. *International Journal of Psychophysiology*, *62*(1), 141–151. https://doi.org/10.1016/j.ijpsycho.2006.03.002
26

**Author contributions**

P.S. developed the study concept. All authors contributed to the study design. A.S.L. performed testing and data collection. A.M., A.S.L. and S.B. performed the data analysis and all the authors interpreted the data. P.S., A.M., A.S.L. drafted the manuscript. All authors approved the final version of the manuscript for submission.

**Competing financial interests**

The authors declare no competing financial interests.

**Acknowledgments**

We would like to thank Dr. Adrienne Wood for kindly providing us with the stimuli used in the present study (see Wood et al., 2016). We also want to thank Olivia Baldi, Flavia Bartolini and Elena Marini for their valuable contribution with the EEG data collection.